\documentclass[reprint, english, amsmath,amssymb,aps,nofootinbib, prd]{revtex4-1}

\usepackage[T1]{fontenc}
\usepackage[latin9]{inputenc}
\usepackage{color}
\usepackage{graphicx}
\usepackage{amssymb,amsmath}
\usepackage{babel}

\def\be{\begin{equation}}
\def\ee{\end{equation}}

\makeatletter

\@ifundefined{definecolor}
{\usepackage{color}}{}

\newcommand{\beq}{\begin{equation}}
\newcommand{\eeq}{\end{equation}}  
\newcommand{\ba}{\begin{eqnarray}}
\newcommand{\ea}{\end{eqnarray}}
\newcommand{\bef}{\begin{figure}}
	\newcommand{\eef}{\end{figure}}

\newcommand{\p}{\partial}

\begin{document}
	
	\title{ Thermalized axion inflation: \\   natural and monomial inflation with small $r$}
	
	\author{Ricardo Z. Ferreira$^{1}$}
	\email{ferreira@icc.ub.edu}
	\author{Alessio Notari$^{1}$}
	\email{notari@fqa.ub.edu}
	
	\affiliation{$^{1}$ Departament de F\'isica Qu\`antica i Astrofis\'ica \& Institut de Ci\`encies del Cosmos (ICCUB), Universitat de Barcelona, Mart\'i i Franqu\`es 1, 08028 Barcelona, Spain}

	\begin{abstract}

		A safe way to reheat the universe, in models of natural and quadratic inflation, is through shift symmetric couplings between the inflaton $\phi$ and the Standard Model (SM), since they do not generate loop corrections to the potential $V(\phi)$. We consider such a coupling to SM gauge fields, of the form $\phi F\tilde{F}/f$, with sub-Planckian $f$. 
		In this case gauge fields can be exponentially produced already {\it during inflation} and thermalize via interactions with charged particles, as pointed out in previous work. This can lead to a plasma of temperature $T$ during inflation and the thermal masses $g T$ of the gauge bosons can equilibrate the system. In addition, inflaton perturbations $\delta \phi$ can also have a thermal spectrum if they have sufficiently large cross sections with the plasma. In this case inflationary predictions are strongly modified: (1) scalar perturbations are thermal, and so enhanced over the vacuum, leading to a generic way to {\it suppress} the tensor-to-scalar ratio $r$; (2) the spectral index is $n_s-1=\eta-4\epsilon$. After presenting the relevant conditions for thermalization, we show that thermalized natural and monomial models of inflation agree with present observations and have $r\approx 10^{-3} - 10^{-2}$, which is within reach of next generation CMB experiments.
	\end{abstract}

	\maketitle

	\section{Introduction}
	
	Cosmology, and in particular the CMB with the latest Planck results~\cite{Ade:2015ava}, entered a precision epoque which allows us to test different classes of models against data. Minimally coupled monomial models of inflation ($V(\phi)\propto \phi^n$, with $n\geq2$)~\cite{Linde:1983gd} are very disfavored and even the well motivated natural inflation scenario~\cite{Freese:1990rb} does not seem to be in the sweetest spot of Planck. 
	
	Monomial models have always received great attention for their simplicity and ease to drive inflation.
	On the other hand, natural inflation remains one of the most appealing scenarios, since its potential, $V(\phi) \approx \Lambda^4 \left[1+\cos(\phi/f_\phi)\right]$, is generated non-perturbatively and is protected against loop corrections by a classical shift symmetry. This is similar to the QCD axion, except that a much higher scale, $\Lambda\gg \Lambda_{QCD}$, is needed, together with $f_{\phi} \gtrsim  M_{Pl}$. 
	
	In both cases $\phi$ needs to be coupled to the Standard Model (SM) in order to reheat the universe after inflation. For natural and quadratic inflation a safe way to reheat is via (classical) shift symmetric derivative interactions, which do not generate new loop corrections to $V(\phi)$. 
	
	For natural inflation the lowest dimensional interactions with the SM, consistent with its symmetries, are 5 dimensional derivative axial couplings, suppressed by a new scale $f$. Importantly, $f$ needs to be sub-Planckian to have successful reheating, otherwise $\phi$ would mostly decay into gravitons, leading to a non-realistic cosmology. When such a coupling is with gauge fields, then gauge bosons can be exponentially produced {\it during} slow-roll inflation~\cite{Tkachev:1986tr, Anber2006} in a significant amount if $f\lesssim 10^{-2} M_P$. In the axion context this is always present: even if we coupled the axion only to SM fermions (or to the Higgs) we would anyway generate a coupling to gauge bosons via loops. Inspired by this, we also consider the effects of the same axial coupling for quadratic, cubic and quartic potentials.
	
	In recent work~\cite{Ferreira:2017lnd} we showed that such a coupling of $\phi$ to SM gauge fields can sustain a thermal bath already {\it during} inflation, with a temperature $T$ larger than the Hubble rate $H$, sourced by the exponential particle production; we called such scenario Thermalized Axion Inflation ({\it ThAI-flation}). In this case, reheating is then simply the time at which the plasma dominates over $V(\phi)$.
	All the dynamics, remarkably, can be generated just by the inflaton axially coupled to SM particles. However, as we will see later, phenomenological requirements oblige us to have some departure from this simplest setup and include shift-symmetric couplings to some SM extension.
	
	The most striking predictions of ThAI-flation are realized when the inflaton perturbations $\delta \phi$ are also thermally distributed,  which can happens if $\phi$ has sufficiently large cross sections with the particles in the plasma. In this case $\delta \phi$ fluctuations are larger than standard vacuum ones. On the other hand, tensors remain in the vacuum since they are much harder to thermalize. Therefore, a generic feature of fully thermalized ThAI-flation ({\it i.e.} with a thermal $\delta\phi$) is a suppression of the tensor-to-scalar ratio $r$, proportional to $H/T$, which would occur to any inflationary potentials. The spectral tilt is also modified to $ n_s-1=\eta-4\epsilon$, where $\epsilon \equiv M_p^2 (V'/V)^2/2$ and $\eta \equiv M_p^2 V''/V$. As we will show, this changes the predictions of natural inflation and makes monomial models agree with present observations, predicting a smaller $r \approx 0.001-0.01$, but still within reach of future experiments~\cite{Delabrouille:2017rct, Matsumura:2013aja, Abazajian:2016yjj}.

	\section{Review of ThAI-flation} We consider the Lagrangian 
	\begin{eqnarray} \label{Lagrangian}
	{\cal L} \supset {\cal L}_{SM} -  \frac{\left(\p_\mu \phi \right)^2}{2}- V(\phi) - \frac{\phi}{4f} F_{\mu \nu} \tilde{F}^{\mu \nu} \, ,
	\end{eqnarray}	
	where $\phi$ is the (axion)-inflaton, $F_{\mu \nu}$ is the strength tensor of some gauge field (in the SM or in some extension of it), $\tilde{F}^{\mu \nu} = \epsilon^{\mu \nu \alpha \beta} F_{\alpha \beta}/(2\sqrt{-g})$ and ${\cal L}_{SM}$ contains the SM fields.
	
	Due to the axial coupling the two circular polarizations of the gauge fields $A_\pm$ satisfy, in a FLRW metric with conformal time $\tau$ and for a homogeneous $\phi(\tau)$, the equation of motion~\cite{Tkachev:1986tr, Anber2006}:
	\begin{eqnarray}
	A_\pm''+\omega^2_\pm A_\pm=0, \,\,\, \,\,\,\, \omega^2_\pm \equiv k^2 \pm \frac{ k \phi'}{ f} \, , \label{eom}
	\end{eqnarray}
	where $'\equiv d/d\tau=a \, d/dt$,  $a$ is the scale factor and $t$ is the cosmic time. A nonzero $\dot{\phi}=d\phi/dt$ creates an instability band at low spatial comoving momenta $k$. During slow-roll $\dot{\phi}$ and $H\equiv \dot{a}/a$ are almost constant and $a\simeq-1/(\tau H)$, so  $\omega^2_\pm=k^2 \mp 2 k \xi/\tau$, where $\xi \equiv \dot{\phi}/(2 f H) \simeq const$. Therefore, one of the two helicities, $A$, is exponentially produced while $ (8\xi)^{-1}  \lesssim |k\tau| \lesssim 2 \xi$ and saturates at $A \propto e^{\pi \xi}$ when $ (8\xi)^{-1}\lesssim |k\tau|$.
	
	This particle production produces large occupation numbers $N_A$ and greatly enhances scattering rates. For example, even if $A$ are photons, the pair production rate of charged particles $A A \leftrightarrow \ell \bar{\ell}$ enters in the Boltzmann equations with a factor $N_A^2$. Since $N_A$ grows exponentially with $\xi$, the SM sector can thermalize through such interactions. Similarly $A A \leftrightarrow \phi \phi$ is enhanced by $N_A^2$; however, this is also suppressed by powers of $1/f$ and so $\phi$ particles have stronger requirements for thermalization. 
	
	As argued in~\cite{Ferreira:2017lnd}, after thermalization $A$ acquires a thermal mass $m_T$, due to loops of thermally distributed charged particles;  for $SU(N)$, for example, $m_T=g T$, with $g=g_1 ^2 (N+N_f/2)/3$, $g_1$ the gauge coupling and $N_f$ the number of flavors. This gives a positive contribution $m_T^2 a^2$ to $\omega_\pm^2$, shielding the instability band and suppressing particle production at low $k$. Moreover, the rate of energy density going into the gauge fields $\dot{\rho}_A$ is proportional to $\int d^3k\, k \, d/dt(|A_+|^2-|A_-|^2)$. This asymmetry is suppressed in the thermal regime, up to differences in $\omega_+-\omega_-$, thus reducing the efficiency of the process. However, the thermal bath cannot dilute completely because, when $T$ decreases, so does $m_T$ and the instability band reopens. For this reason the system should reach an equilibrium\footnote{Or perhaps with oscillations around the equilibrium point.} when the two effects balance each other. This happens if $m_T$ equals the negative contribution $-\xi H$ in $\omega_\pm$, leading to the equilibrium temperature~\cite{Ferreira:2017lnd} 
	\begin{eqnarray} \label{temperature}
	T_{eq}= \frac{\xi}{g} H \, .
	\end{eqnarray}
	When this equilibrium stage is reached, the properties of the system change dramatically from an exponential to a linear dependence on $\xi$, leading to new predictions. For this reason the constraints~\cite{Barnaby:2011vw,Ferreira2014a, Ferreira2015a,Peloso:2016gqs} on $\xi$ are very alleviated and much larger values of $\xi$ are allowed, as explained in~\cite{Ferreira:2017lnd}. Note also that in this letter we will not consider the regime where  $A$ backreacts onto $\phi$, which has very non-trivial friction effects~\cite{Anber2009, Notari2016}.
	
	If the gauge field belongs to the SM, this dynamics thermalizes the SM already during inflation. However,  observationally the most interesting regime is when $\phi$ perturbations are also thermalized. This means that on sub-horizon scales they follow the Bose-Einstein distribution $N_\text{BE}$. Note also that an important feature of this scenario is that the inflaton cannot receive thermal mass contributions, as a consequence of the shift symmetry of the interactions\footnote{Temperature effects could affect the axion propagator leading to a renormalization of the field strength and perhaps also to an effective friction for the perturbations \cite{Anber2009, Eichhorn:2012uv}. However, since we work in non-backreacting regime we expect such effective friction to be small.}. Evaluating $N_\text{BE}$ at horizon crossing, when the scalar curvature perturbation $\zeta$ freezes out, leads to the scalar power spectrum \cite{Berera1995, Ferreira:2017lnd},
	\begin{eqnarray} \label{Power spectrum}
	P_\zeta =2  P_\zeta^\text{vac}  \left[\frac{1}{2}+N_\text{BE}\right]_{\frac{k}{a}=H}  = P_\zeta^\text{vac} \left(1+ \frac{ 2T_{eq}}{H}\right), \nonumber
	\end{eqnarray}
	where $P_\zeta^\text{vac} \equiv H^4/(2\pi \dot{\phi})^2$. The first term corresponds to the standard vacuum result and the second term are the additional thermal excitations. On the other hand, the tilt of the spectrum receives new contributions from the time variation of $T/H$, which is controlled by $\dot{\phi}$ and $H$, so that we get a new function of the slow-roll parameters \cite{Ferreira:2017lnd}
	\begin{eqnarray} \label{tilt}
	\frac{d \log(P_\zeta)}{d \log(k)}=n_s-1 = \eta -4\epsilon.
	\end{eqnarray}
	As we will see, for the potentials considered here, this implies a slightly bluer spectrum. 
	
	The tensors instead are likely to remain in vacuum, since gravitational couplings are too weak to thermalize them. This fact, combined with the enhancement of $P_\zeta$, implies that $r$ is suppressed \cite{Berera1995, Ferreira:2017lnd}
	\begin{eqnarray}
	r= \frac{r_\text{vac}}{1+ 2 T_{eq}/H}, \quad \quad r_\text{vac} \equiv 16 \epsilon \, .
	\label{r}
	\end{eqnarray}
	These two generic features, larger $n_s$ and smaller $r$, are interesting as they change natural inflation predictions and they bring polynomial models of inflation back in agreement with data, as we will discuss in section~\ref{3}.
	
	\subsection{ Thermalization of inflaton perturbations} In order to thermalize inflaton perturbations $\delta\phi$ during the equilibrium stage the thermal cross sections should be large enough. In~\cite{Ferreira:2017lnd} we estimated this in the minimal case of SM gauge fields, even Abelian, interacting via $A A \leftrightarrow \phi \phi$. 
	There can be, however, more efficient interactions involving $\phi$. For example, if $A$ is non-Abelian the cross sections $A A \leftrightarrow \phi A$ exist, due to self-couplings, and have less powers of $1/f$. In some axion models an even larger rate comes from a derivative coupling with the top quark ($t$) of the form \cite{Dine:1981rt, Zhitnitsky:1980tq, Salvio:2013iaa}
	\begin{eqnarray}
	{\cal L}_\text{tot}  \supset  \frac{\p_\mu \phi}{\tilde{f}} \sum_{X=L,R} c_X \bar{t}_X \gamma_\mu  t_X \, , \label{a-top-Higgs}
	\end{eqnarray}
	where $t_X$ are Weyl fermions, $\tilde{f} \equiv \alpha_s f/(2\pi)$, $\alpha_s\equiv g_1^2/(8\pi)$  and $c_X$ are numbers of ${\cal O}(1)$. 
	Because the coupling is already present at tree level it is stronger, by a factor $\alpha_s/2\pi$, than the axial coupling between $A$ and $\phi$, which is generated at loop level. 
	Thus, the processes $t \bar{t} \leftrightarrow \phi h$ and $t \phi \leftrightarrow t h$ can be large due to the sizable top Yukawa coupling, $y_t$, with the Higgs ($h$), leading to a total production cross section\footnote{Here $\alpha_s$ appears in the denominator just due to our convention for $f$.} \cite{Salvio:2013iaa}: 
	\begin{eqnarray} \label{cross-section}
	\sigma_{t}=\frac{37 (c'_t)^2 y_t^2}{8 f^2 \pi \alpha_s^2} \, ,
	\end{eqnarray}
	where $c'_t=c_R -c_L$, and we assumed no direct coupling between $\phi$ and $h$.
	At energies of around $10^{16}$ GeV and for $c'_t \simeq 1$  this cross section is larger than $\sigma_{A \,A \leftrightarrow \phi \, A}$, with $A$ being gluons, by a factor ${\cal{O}}(10^3)$.
	The condition to have $\delta\phi$ thermalized is thus $n_{h,t} \,\sigma_t \gg H $ which requires
	\begin{eqnarray} \label{Thermalization condition}
	\left(\frac{f}{H}\right)^2 \ll  \frac{37}{8\pi}  \frac{c_t y_t \xi^3}{\alpha_s^2 g^3} \, , \label{therm2} 
	\end{eqnarray}
	where we used eq.~(\ref{temperature}) and the number density $n_{h,t} \simeq T^3/\pi^2$.
	
	There is, however, another constraint: ensure that $T_{eq}$ is below the UV cutoff,
	$\tilde{f}$.
	%= \frac{\alpha_s f}{2\pi}. \label{cutoff}
	If $T_{eq}>\tilde{f}$ this signals the need of considering new dynamical degrees of freedom ({\it i.e.} fermions that run in a triangle loop, generating the axial coupling), and one should check if the instability in $A$ can still be present for momenta around $\xi H=g T$. This constraint turns out to be quite stringent and stronger than the no-backreaction condition. 
	
	In the scenario with only $\phi$, the SM and the axial coupling, the UV constraint cannot be satisfied after imposing observational constraints. Thus, we need to depart from this minimal scenario by enhancing the cross sections involving $\phi$ and the particles in the plasma. This could be achieved, {\it e.g.}, with  $c'_T \gtrsim 1$, several Higgs-like particles with similar Yukawa couplings to $t$, several colorless fermions also with similar Yukawa couplings to $h$ or, instead, a larger axion-gluon cross section by considering a larger gauge group. We keep using the interaction in eq.~(\ref{cross-section}) for keeping $\delta\phi$ thermal during inflation, but enhanced by one of the above options, which we parameterize by a larger $c_t'$.
	
	Finally, even if $t$ and $\phi$ are in equilibrium we also need to ensure that $A$, which suffers the instability and sustains the process, is kept in thermal equilibrium at $T=\xi/g$. To achieve this, the most efficient way is by self-interactions. For $SU(3)$ the typical cross section is
	\begin{eqnarray}
	\sigma_{AA\leftrightarrow AA} \simeq \frac{9 \pi \alpha_s^2}{8 T^2}
	\end{eqnarray}
	However, the phenomenological requirements are not met with the SM value of $\alpha_s$ at inflationary scales, $\alpha_s \simeq 1/40$. For this reason we need to consider an enhanced value~\footnote{This can be achieved with a new gauge group at high energy with either a larger coupling, larger $N$ or with extra charged particles at high energies affecting the running of $\alpha_s$.} of $\alpha_s \simeq 1$.
	
	Clearly these last two requirements introduce some model dependence. However, this model dependence is only reflected in the precise value of $r$, and not in $n_s$. 
	
	\subsection{ Reheating}
	
	Another interesting feature of ThAI-flation is its natural reheating process. The $\phi$ field can transfer all its energy to the plasma, through three different channels: (1) scatterings, with rate $\Gamma_s$, (2) parametric resonance~\cite{Traschen:1990sw, Kofman:1997yn, Adshead2015}  in an oscillating field $\phi=\bar{\phi}(t) \sin(m t)$, with~\footnote{This can be obtained by writing eq.~(\ref{eom}), as a Mathieu equation, $d^2 A/dz^2 + (4k^2/m^2 - 2 q \cos(2z)) A=0$, where $ z\equiv mt/2$ and $q\equiv 2k \bar{\phi}/(m f)$ is the band width calculated in the first resonance, $k\sim m/2$, and we took into account the redshift of the modes that suppresses the resonance.} $\Gamma_r \approx q^2 m = m \phi^2/(2 f^2)$, (3) perturbative decay, with $\Gamma_d\approx m^3/(64\pi f^2)$. 
	
	In the realization we consider here the first channel dominates. In fact, when inflation ends $m \gtrsim H$, the zero mode $\phi(t)$ behaves as matter and can participate in scattering events. The leading process to thermalize is again through the interaction eq.~(\ref{a-top-Higgs}). In this case the $\phi$ particle number satisfies the Boltzmann equation
	\begin{eqnarray}
	\frac{1}{a^3} \frac{d (a^3 N_\phi)}{dt} \approx - \sigma_{t} \, n_{h,t} N_\phi.
	\end{eqnarray}
	
	Thermalizing $\phi$ at the end of inflation then corresponds to the same condition as in eq.~\ref{a-top-Higgs}, which means that reheating is almost instantaneous. 
	
	\section{ Results \label{3}}
	Now we are able to explore the predictions for several inflationary potentials in the thermal regime. We consider {\it natural inflation}, $V(\phi) =  \Lambda^4 \left[1+ \cos (\phi/f_\phi) \right]$, \cite{Freese:1990rb} and monomial potentials, $V(\phi) \propto  \phi^n$ \cite{Linde:1983gd}, for $n=2,3,4$.
	We distinguish the scale $f_\phi$, which needs to be super-Planckian in order to be compatible with observations, from $f$, the coupling to SM fields, which needs to be sub-Planckian to reheat the universe and trigger thermalization.
	
\begin{figure}
		\includegraphics[scale=0.5]{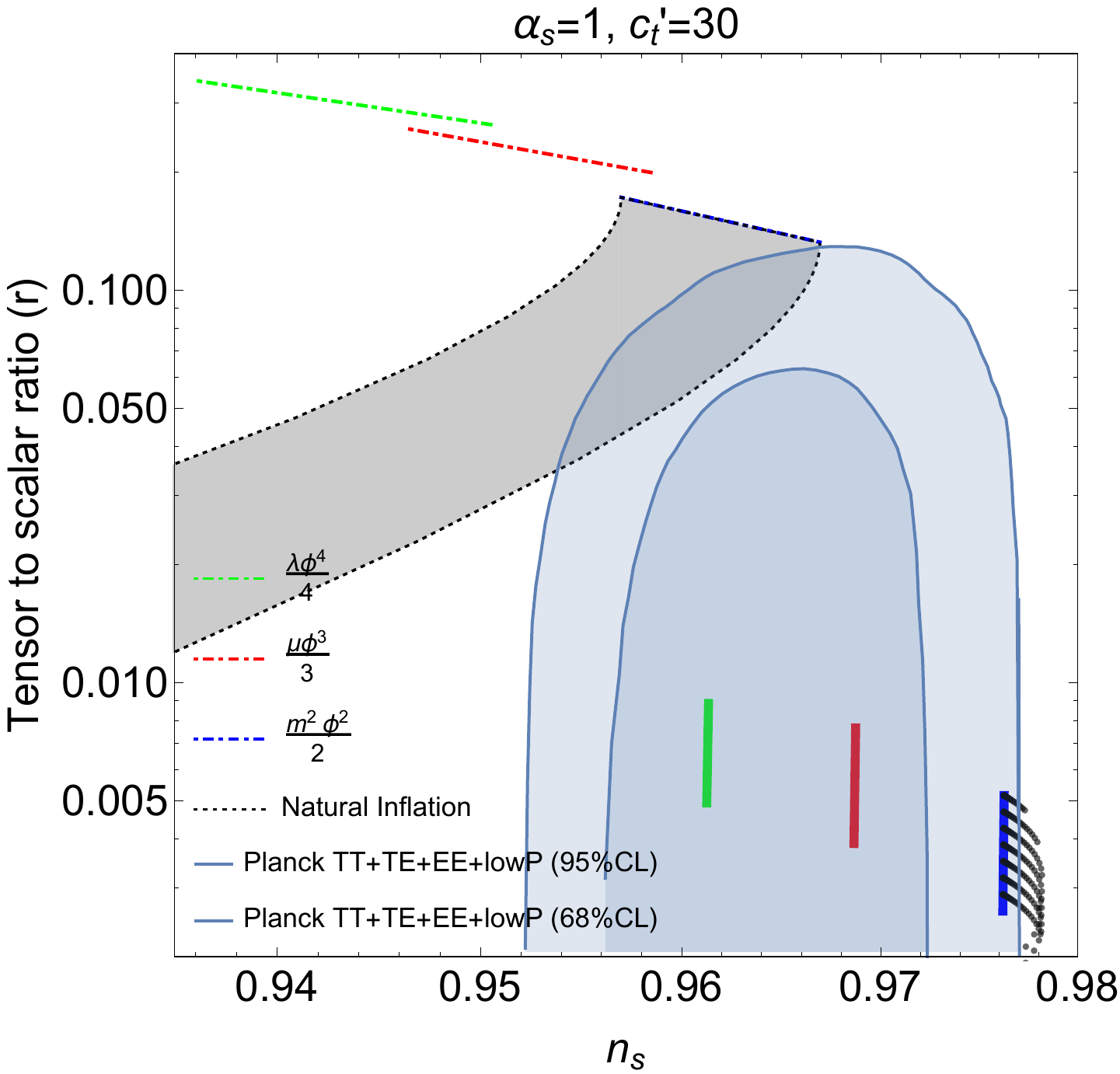}
		\caption{Predictions of each potential in the $n_s-r$ plane. Dot-dashed lines are the vacuum predictions (from 46 to 60 e-folds) while solid lines are predictions in the thermalized regime, which have no uncertainty on the e-folds, due to instantaneous reheating.  The solid lines correspond to values of $f$ between $8 \times 10^{-4} \lesssim f/M_p \lesssim 2 \times 10^{-3}$. In turn this corresponds to $2 \lesssim 10^{15}  \lambda \lesssim 4$ for quartic inflation, $5 \lesssim 10^{14}   \mu/M_p \lesssim 10$ for cubic inflation, $8 \lesssim 10^{7}  m/M_p \lesssim 10$ for quadratic inflation and $2 \lesssim 10^3 \Lambda/M_p \lesssim 20 $ and $5\lesssim f_\phi/M_p\lesssim 10^3$ for natural inflation (saturating to the blue solid line for larger values of $f_\phi$).}
		\label{ns-r plot}
	\end{figure}

	We fix one parameter of the model by requiring the amplitude of the power spectrum in eq.~(\ref{Power spectrum}) to match observations, {\it i.e}, $P_\zeta = 2.2 \times 10^{-9}$ \cite{Ade:2015ava}. Then, we compute the number of e-folds of a given scenario using instantaneous reheating, as follows from the previous section. 
	The values of $f$ and $f_\phi$ are chosen such that during inflation the parameters which control thermalization, $f/H$ and $\xi$, lie in a region where $\phi$ is thermalized and $T_{eq}$ is below the UV cutoff which corresponds to $8  \lesssim 10^{4} f/M_p \lesssim 20$ and $5\lesssim f_\phi/M_p$.
	\begin{figure}
		\includegraphics[scale=0.5]{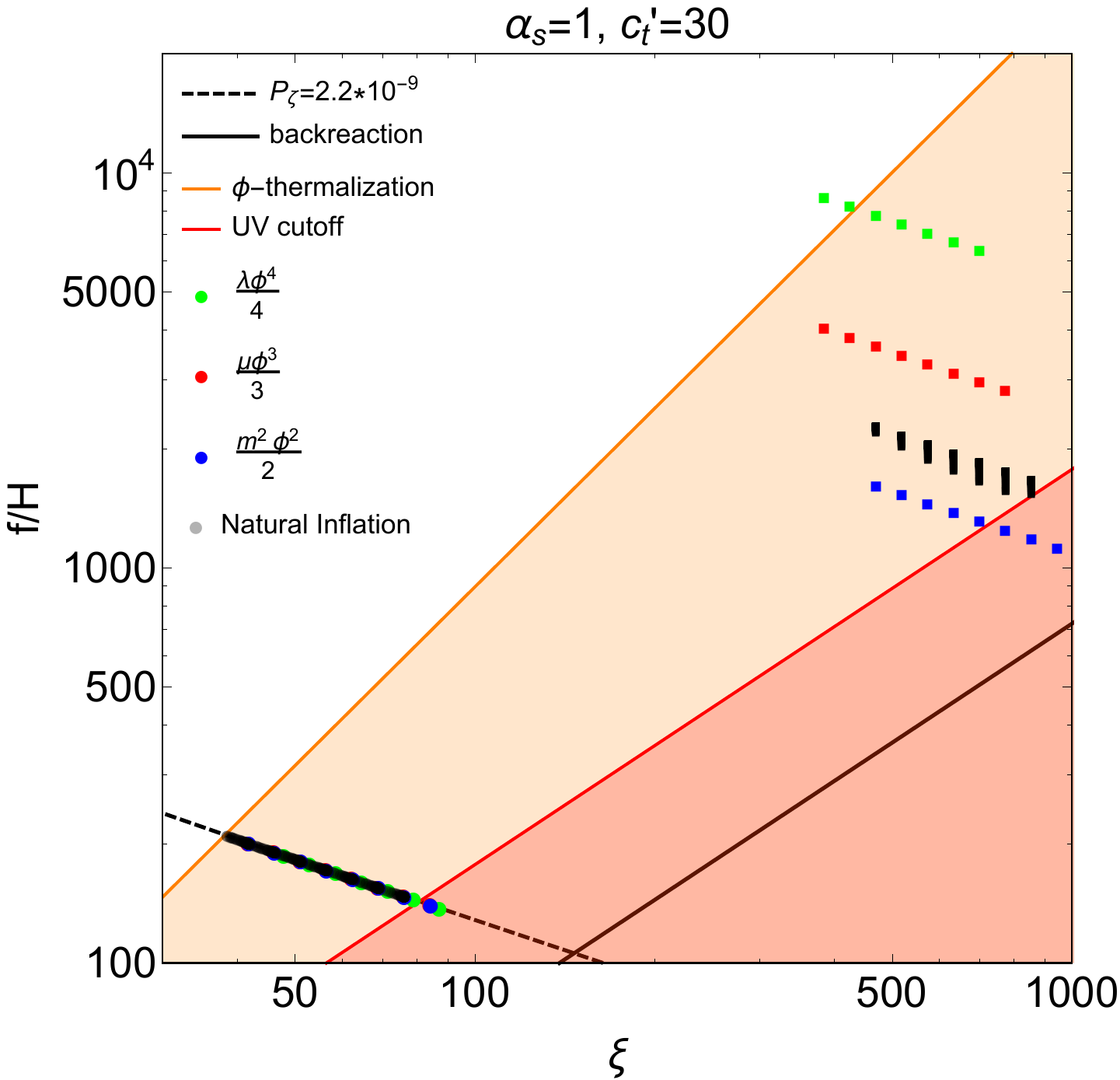}
		\caption{Evolution, for each potential, in the $f/H$ vs $\xi$ plane. Circles correspond to CMB scales exiting the horizon and squares to end of inflation. The orange region corresponds to thermalization of $\phi$, while the red corresponds to $T_{eq}$ above the UV cutoff. Finally the black dashed line is the observational constraint on $P_\zeta$ and below the black solid line there is the backreacting region.}
		\label{f/H-xi plot}
	\end{figure}
	In figs.~\ref{f/H-xi plot} and~\ref{ns-r plot} we fix $c'_t=30$, $\alpha_s = 1$ and a Yukawa coupling $y_t = 0.5$. Larger values of $c'_t$ would allow for a larger parameter space and, in particular, larger $f$ and smaller $\xi$. From eqs.~(\ref{temperature}) and~(\ref{r}) this would correspond to a larger $r$. In the setup chosen here the thermalized system is composed of gluons, $t$, $\phi$, $h$ and possibly the additional particles which effectively increase $c'_t$ and $\alpha_s$. We consider these extra particles to be color singlets, such that the $g$ entering in $m_T$ remains unaffected (and we use $N=3$). The no-backreaction condition does care about the total number of particles in the plasma but this condition is still weaker than the UV cutoff constraint.
	
	In fig.~\ref{f/H-xi plot} we can see the evolution in the $f/H$ vs. $\xi$ plane. Different colors represent different potentials. Circles correspond to the time at which CMB scales exited the horizon and squares correspond to the end of inflation. Initially all points are on top of the dashed black line to match the right amplitude of the spectrum at CMB scales.
	
	In fig.~\ref{ns-r plot} we show the associated predictions in the $n_s-r$ plane. It is immediate to see that for all models considered the spectrum becomes bluer than the standard case, as a consequence of eq.~(\ref{tilt}) and $r$ is reduced by ${\cal O}(10-100)$. Note also that the instantaneous nature of reheating eliminates uncertainties on the total number of efolds and so leads to a sharp prediction on $n_s$. This behavior brings monomial models in agreement with data. Regarding natural inflation the predictions are also changed and the region with $f_\phi<16M_p$ is excluded by the $68\%$ data contour~\footnote{This makes the issue of superplanckian field excursions more relevant than in the standard case, for a discussion see~\cite{Banks:2003sx,Hebecker:2016dsw}.}. Moreover, for each potential there is a lower and upper bound on $r$ which can be traced back to the region of $\xi$ in fig.~\ref{f/H-xi plot} where $\phi$ is thermalized and below the UV cutoff. Although the value of $r$ has some model dependence, in the number of extra particles we needed to consider and the specific interaction which thermalizes $\delta\phi$, the values of $n_s-1$ are robust and so can be seen as accurate predictions for models of ThAI-flation with thermal $\delta \phi$.
	
	\section{Conclusions}
	
	In this letter we explored the predictions of ThAI-flation. Axial couplings with gauge fields, which have a classical shift symmetry, are used to reheat the universe without spoiling the inflationary potential, and as a by-product this can lead to a thermal bath during inflation. The bath is not diluted because it is constantly sourced by the instability in the gauge fields generated by such coupling. We considered the case in which interactions are strong enough to also keep the inflaton perturbations in thermal equilibrium, while keeping the temperature below the UV cutoff. 
	
	The outcome of this mechanism is that  scalar perturbations are thermally enhanced, while  tensors are unaffected. This leads a suppression of the tensor-to-scalar ratio $r$, as anticipated in \cite{Ferreira:2017lnd}, which can be used to rescue models of inflation, which are otherwise ruled out, such as monomial models.
	
	In this scenario all the thermal dynamics comes from one extra parameter, the axial coupling $f$, which determines both the spectrum of perturbations and the reheating history, provided the inflaton is kept in thermal equilibrium. 
	An efficient way to do this is via shift-symmetric couplings with fermions, such as the top quark, due to its large Yukawa coupling.  However, in order to satisfy phenomenological constraints we needed to consider additional particles beyond the SM with stronger couplings. This introduces some model dependence in the predictions on $r$, but not on the spectral tilt $n_s$. The latter is determined sharply, due to the instantaneous reheating of such models.
	
	We derived then the predictions of ThAI-flation for $n_s$  and $r$ for natural inflation and monomial potentials. For all these scenarios $n_s$ becomes bluer. For natural inflation the predictions change significantly. For monomial potentials this leads to agreement with present data.  In all cases, and for the setup studied here, $r\approx10^{-3}-10^{-2}$ which is within reach of future observations.
	
	For the future a dedicated analysis of non-Gaussianities needs to be performed to ensure the full viability of this scenario.

	${}$\linebreak
	\emph{\textbf{Acknowledgments:}}
	We thank Federico Mescia, Luca di Luzio and Paolo Panci for useful discussions. This work is supported by the grants EC FPA2010-20807-C02-02, AGAUR 2009-SGR-168, ERC Starting Grant HoloLHC-306605 and by the Spanish MINECO under MDM-2014-0369 of ICCUB (Unidad de Excelencia ''Maria de Maeztu'').

	\bibliographystyle{JHEP}
	\bibliography{ThAIPredictions}
	
	%%%%
\end{document}